\documentclass[twocolumn,amsmath,amssymb,nofootinbib,superscriptaddress,longbibliography]{revtex4-1}

\usepackage{graphicx}
\usepackage{amsmath}
\usepackage{amssymb}
\usepackage{xcolor}
\usepackage{hyperref}
\hypersetup{pdfstartview={FitH},pdfpagemode={UseNone}, breaklinks=true, colorlinks=true,bookmarks=false,linkcolor=blue, citecolor=blue, urlcolor=blue, bookmarksopen=true, pdfnewwindow=true}
\usepackage[all]{hypcap}
\usepackage[english]{babel}
\usepackage{physics}

\newcommand{\eps}{\varepsilon}

\begin{document}

\preprint{APS/123-QED}

\title{Electric circuit emulation of topological transitions driven by quantum statistics}

\author{Nikita A. Olekhno}
\thanks{These two authors contributed equally}
%\email{nikita.olekhno@metalab.ifmo.ru}
\affiliation{School of Physics and Engineering, ITMO University, Saint Petersburg 197101, Russia}
\author{Alina D. Rozenblit}
\thanks{These two authors contributed equally}
\affiliation{School of Physics and Engineering, ITMO University, Saint Petersburg 197101, Russia}
\thanks{These two authors contributed equally}
\author{Alexey A. Dmitriev}
\affiliation{School of Physics and Engineering, ITMO University, Saint Petersburg 197101, Russia}
\author{Daniel A. Bobylev}
\affiliation{School of Physics and Engineering, ITMO University, Saint Petersburg 197101, Russia}
\author{Maxim A. Gorlach}
 \email{m.gorlach@metalab.ifmo.ru}
\affiliation{School of Physics and Engineering, ITMO University, Saint Petersburg 197101, Russia}

\date{\today}

%_________________________Abstract_________________________

\begin{abstract}

Topological phases exhibit a plethora of striking phenomena including disorder-robust localization and propagation of waves of various nature. Of special interest are the transitions between the different topological phases which are typically controlled by the external parameters. In contrast, in this Letter, we predict the topological transition in the two-particle interacting system driven by the particles' quantum statistics. As a toy model, we investigate an extended one-dimensional Hubbard model with two anyonic excitations obeying fractional quantum statistics in-between bosons and fermions. As we demonstrate, the interplay of two-particle interactions and tunneling processes enables topological edge states of anyon pairs whose existence and localization at one or another edge of the one-dimensional system is governed by the quantum statistics of particles. Since a direct realization of the proposed system is challenging, we develop a rigorous method to emulate the eigenmodes and eigenenergies of anyon pairs with resonant electric circuits.
\end{abstract}

\maketitle

%_______________________Introduction_______________________
{\it Introduction.~}---~Anyons are quantum quasi-particles with the statistics intermediate between bosons and fermions that can exist only in one- and two-dimensional systems~\cite{1982_Wilczek, 1977_Leinaas}. Successfully applied to the theoretical description of fractional quantum Hall effect~\cite{1984_Arovas, 1984_Halperin}, they later attracted considerable attention as promising candidates to realize fault-tolerant quantum computations~\cite{2008_Nayak, 2013_Stern}. However, despite realizations of fractional quantum Hall physics in two-dimensional condensed matter systems ranging from thin-film semiconductor nanostructures~\cite{1982_Tsui, 2008_Dolev} to graphene~\cite{2009_Bolotin}, a direct experimental investigation of anyons remains extremely challenging with a few studies currently available~\cite{2020_Bartolomei, 2020_Nakamura}. To overcome the complexity of experiments with real quasi-particles obeying anyonic statistics, their analog emulators based on bosonic~\cite{2011_Keilmann} and, particularly, photonic systems~\cite{2008_Cho, 2012_Longhi, 2020_Noh}, magnetic metamaterials~\cite{2018_Todoric}, and even classical mechanical setups~\cite{2020_Fruchart} were put forward.

Currently, the research interest is shifting from the studies of individual anyons to the states of multiple quasi-particles exhibiting rich physics such as topological phases~\cite{2008_Nayak}. Over the last decade, the concepts of topology attained an increased attention due to the prospect of topologically protected edge and corner states immune to defects and imperfections~\cite{2019_Ozawa} available both in linear and nonlinear~\cite{2020_Smirnova} domains. Moreover, recent studies suggest that the topological protection also applies to the entangled states of quantum light~\cite{2018_Blanco_Redondo, 2018_Mittal,2020_Blanco}.

Two-particle systems described by the Hubbard model~\cite{1986_Mattis} provide an ideal playground to test the interplay of interactions and topology not only through the theoretical analysis~\cite{2016_Di_Liberto,2017_Gorlach,2017_Bello,2017_Marques,2018_Salerno,2019_Zurita}, but also via analog emulators~\cite{2016_Mukherjee,2020_Olekhno} and experiments with cold atoms~\cite{2006_Winkler} or superconducting qubits~\cite{2021_Kim,2021_Besedin}. While two-particle physics of bosons and fermions is relatively well-studied, the physics of a pair of interacting anyons~\cite{2012_Longhi} and especially their topological phases remain largely uncharted.

In this Letter, we predict the topological transition induced by the quantum statistics of interacting particles. Since the direct experimental implementation of our model remains currently out of reach, we design an analog emulator of the proposed system based on electrical circuits. Such a platform~\cite{2015_Ningyuan} has previously been exploited to emulate topological transitions in a range of exotic systems, including nonlinear~\cite{2018_Hadad}, non-Hermitian~\cite{2020_Helbig, 2020_Liu, 2021_Kotwal}, non-Abelian~\cite{2020_Song}, four-dimensional~\cite{2020_Wang}, fracton~\cite{2019_Pretko} and quantum two-boson~\cite{2020_Olekhno} models. Here, we further generalize this approach opening an avenue to emulate statistics-induced topological transitions.

\begin{figure}[t]
    \centering
    \includegraphics[width=8.5cm]{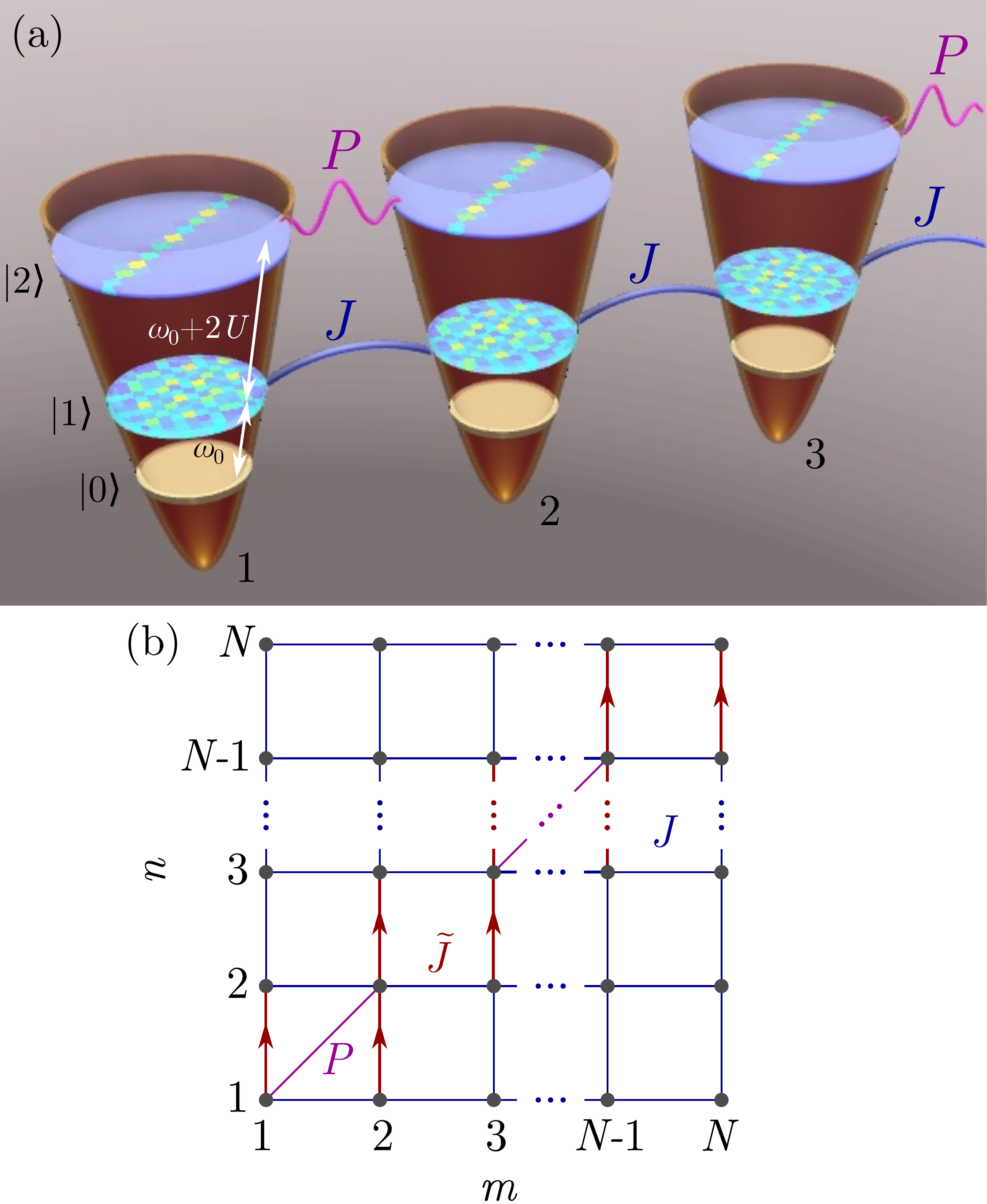}
    \caption{(a) A scheme of the one-dimensional array of $N$ cavities with a pair of anyons. Neighboring cavities are coupled via links $J$ describing the tunneling of one anyon between the adjacent cavities, and links $P$ which correspond to the simultaneous tunneling of both particles; $\omega_{\rm 0}$ is the energy of one anyon, and $2U$ is the energy of their effective on-site interaction. The insets in each cavity illustrate the typical probability distributions for anyon pair. (b) Equivalent two-dimensional tight-binding model with tunneling links $J$, $P$, and complex couplings $\tilde{J}$ marked by the arrow.}
    \label{fig:Model}
\end{figure}

%_______________________Theoretical_model______________________
{\it Theoretical model.~}---~As a specific model, we consider a one-dimensional array of cavities supporting two excitations with anyonic statistics that can tunnel between the adjacent cavities [Fig.~\ref{fig:Model}(a)]. This system is described by the extended Hubbard model:
\begin{multline}
    \hat{H}= \omega_{0}\sum_{m=1}^{N} \hat{n}_{m} + U\,\sum_{m=1}^{N} \hat{n}_{m}(\hat{n}_{m}-1) - J\sum_{m=1}^{N-1} (\hat{a}^{\dagger}_{m} \hat{a}_{m+1} +\\
    + {\rm H.c.}) + \dfrac{P}{2} \sum_{m=1}^{(N-1)/2} (\hat{a}^{\dagger}_{2m-1} \hat{a}^{\dagger}_{2m-1} \hat{a}_{2m} \hat{a}_{2m} + {\rm H.c.}),
\label{eq:Hamiltonian}
\end{multline}
where we set $\hbar =1$ for simplicity. Here, $N$ is the number of cavities, $\omega_{\rm 0}$ is the energy of a single anyon in the cavity, $U$ is the energy of on-site interaction between two anyons, $\hat{a}^{\dagger}_{m}$ and $\hat{a}_{m}$ are creation and annihilation operators for the anyon at the $m$-th cavity, and $\hat{n}=\hat{a}^{\dagger}_{m} \hat{a}_{m}$ is a quasi-particle number operator. The first term in Eq.(\ref{eq:Hamiltonian}) leads to a trivial energy shift by $2\,\omega_{\rm 0}$, while the second term describes the interaction energy of $2U$ for two anyons sharing the same cavity. Finally, amplitudes $J$ and $P$ describe the conventional single-particle tunneling and simultaneous hopping of the two particles between the respective sites. Physically, such model corresponds to the interacting Bose gas~\cite{1999_Kundu}, and the concept of anyonic quasi-particles provides an efficient tool to treat this kind of problem.

%_________________________Figure_1__________________________
%_________________________Figure_1__________________________

In the case of boson pairs, the term $\propto P$ leads to the emergence of interaction-induced topological states of bound boson  pairs~\cite{2020_Olekhno,2020_Stepanenko-PRA}. The difference from the two-boson model is rooted to the commutation relations of anyon creation and annihilation operators~\cite{1999_Kundu,2006_Batchelor}:
\begin{align}
  \hat{a}_{l}\hat{a}_{k}&=\exp(i\theta\, {\rm sgn}(l-k))\,\hat{a}_{k}\hat{a}_{l},\\
   \hat{a}_{l}\hat{a}^{\dagger}_{k}&=\delta_{lk} + \exp(-i \theta\, {\rm sgn}(l-k))\,\hat{a}^{\dagger}_{k}\hat{a}_{l},
  \label{eq:Commutators}
\end{align}
where $\theta$ is the statistical exchange angle, $\delta_{lk}$ is the Kronecker symbol, and ${\rm sgn}(x)$ function is equal to 1 for $x>0$, -1 for $x<0$, and 0 for $x=0$. Limiting cases of $\theta=0$ and $\theta=\pi$ correspond to bosons or pseudo-fermions, respectively. Such  pseudo-fermions satisfy fermionic commutation relations when the particles are located at the different sites and boson commutation relations when the particles share the same cavity.

As the Hamiltonian Eq.(\ref{eq:Hamiltonian}) preserves the number of particles (see Supplementary Materials, Sec.~I), the two-particle wave function $\ket{\psi}$ can be searched in the form $\ket{\psi} = \dfrac{1}{\sqrt{2}} \sum_{m,n=1}^{N} \beta_{mn} \hat{a}^{\dagger}_{m}\hat{a}^{\dagger}_{n}\ket{0}$,
where $\ket{0}$ is the vacuum state and coefficients $\beta_{mn}$ characterize the weight of the state with the first anyon located at cavity $m$ and the other one located at cavity $n$. Due to anyonic statistics, these coefficients are related to each other as $\beta_{mn}=\exp\left(-i\theta\,{\rm sgn}(m-n)\right)\,\beta_{nm}$.

From now on, we focus on the eigenmodes of our model, obtained as the solutions to the stationary Sch{\"o}dinger equation $\hat{H}\ket{\psi}=E\,\ket{\psi}$, where the Hamiltonian $\hat{H}$ is given by Eq.~(\ref{eq:Hamiltonian}) and $E$ is the eigenstate energy. This yields the set of linear equations for the coefficients $\beta_{mn}$ (without loss of generality, we consider $m\geq n$ and set $J=1$):
\begin{small}
\begin{gather}
\beta_{m-1,n} + \beta_{m+1,n} + \beta_{m,n-1} + \beta_{m,n+1} = -\varepsilon\beta_{mn},\mspace{4mu} m-n\geq 2,\label{eq:Tight_binding-1}\\
\beta_{nn}+\beta_{n+2,n}+\beta_{n+1,n-1}+{\rm e}^{-i\theta}\,\beta_{n+1,n+1}=-\eps\,\beta_{n+1,n},\label{eq:Tight_binding-2}\\
\beta_{n-1,n}+\beta_{n+1,n}+e^{i\theta}\,\beta_{n,n-1}+{\rm e}^{-i\theta}\,\beta_{n,n+1}+P\,\beta_{n'n'}\notag\\
=(2U-\eps)\,\beta_{nn},\label{eq:Tight_binding-3}
\end{gather}
\end{small}
where $n'=n\pm 1$ for odd and even $n$, respectively, $\varepsilon=E-2\omega_{0}$ is the mode energy, and $2\,\omega_{0}$ is taken as an energy reference. Further details of the tight-binding analysis are specified in Supplementary Materials, Sec.~I.

It should be stressed that Eqs.~\eqref{eq:Tight_binding-1}-\eqref{eq:Tight_binding-3} can be interpreted as an eigenvalue problem for a two-dimensional (2D) tight-binding model with the sites connected by the links $J$, $J\,{\rm e}^{\pm i\theta}$, and $P$ [Fig.~\ref{fig:Model}(b)]. In this interpretation, the coefficients $\beta_{mn}$ correspond to the amplitudes at the sites of a 2D lattice. Such an approach to unfold the Hilbert space and reformulate two-particle 1D model as a single-particle 2D model~\cite{2013_Longhi} has been applied for bosonic case in a series of proposals~\cite{2012_Longhi,2016_Di_Liberto,2017_Gorlach,2018_Gorlach, 2020_Olekhno}. However, to account for anyonic statistics, complex tunneling links  are required [Fig.~\ref{fig:Model}(b)]. While keeping the system Hermitian, such couplings render some off-diagonal elements of the Hamiltonian $H_{mn,m'n'}$ complex, namely
\begin{equation}
    H_{mn,m'n'}=-J\,{\rm e}^{-i\theta}\:,\mspace{6mu} H_{m'n',mn}=-J\,{\rm e}^{i\theta}\:,
\end{equation}
where the direction from the site $(m,n)$ to the site $(m',n')$ is shown by the arrow in Fig.~\ref{fig:Model}(b).

%_________________________Figure_2__________________________
\begin{figure}[ht!]
    \centering
    \includegraphics[width=\linewidth]{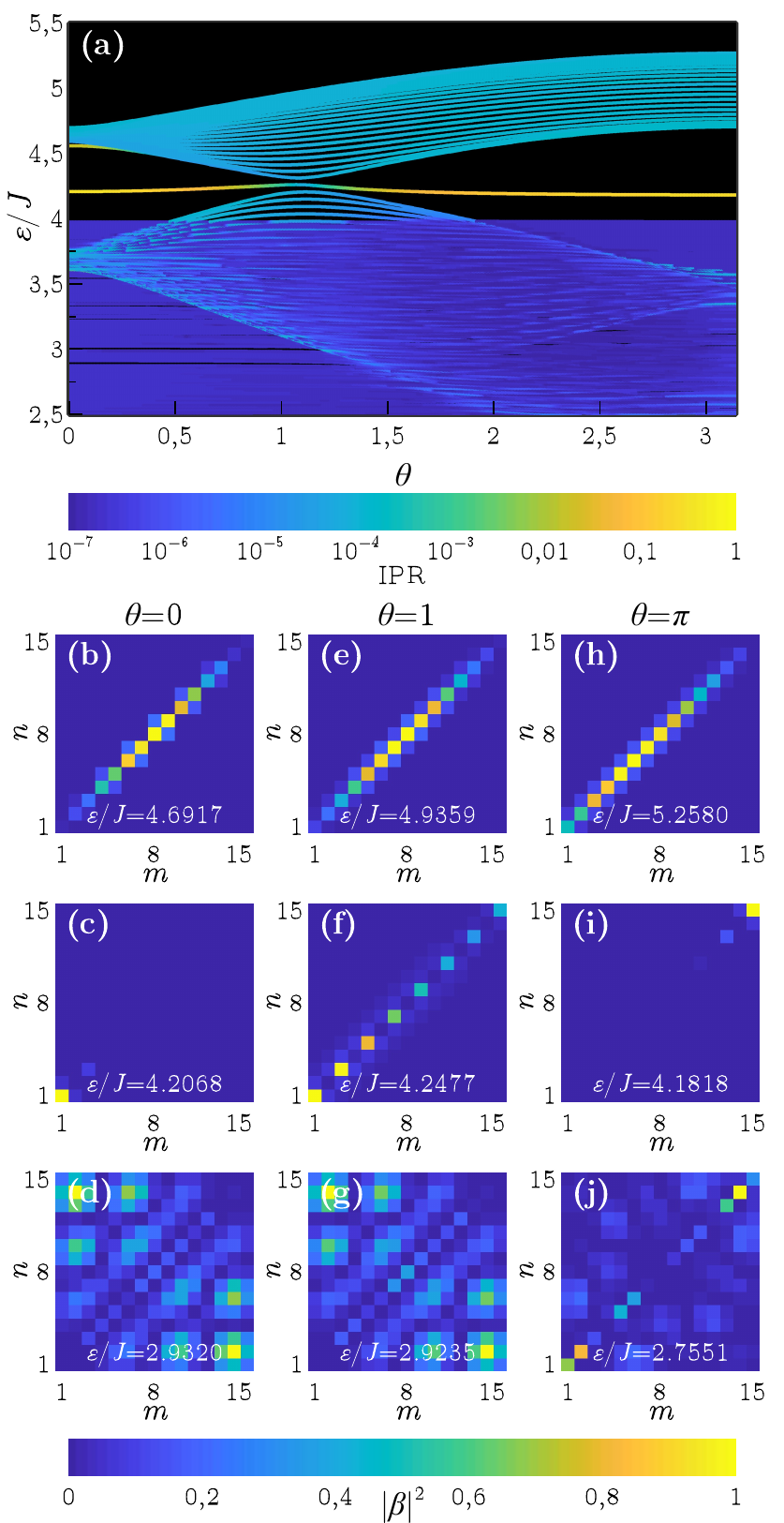}
    \caption{(a) Localization map of the eigenstates ${\rm IPR}(\varepsilon,\theta)$ calculated for $\theta \in [0,\pi]$. The system parameters are $N=45$, $U/J=1.5$, and $P/J=-0.75$. (b-j) Eigenmode profiles $|\beta_{mn}|^{2}$ in the system with $N=15$ sites for the scattering states (d,g,j), bulk doublons (b,e,f,h), and doublon edge states (c,i) calculated for $\theta=0$, $1$, and $\pi$, respectively. Mode energies $\varepsilon$ and values of $\theta$ are shown in the panels.}
    \label{fig:Theory}
\end{figure}
%_________________________Figure_2__________________________

%____________________Topological_edge_states___________________
{\it Statistics-induced topological transition.~}---~The key feature of the outlined model is the topological transition happening with the variation of the angle $\theta$ responsible for the type of the quantum statistics. To grasp the underlying physics, we first analyze strong interaction limit $|U/J|\gg 1$. In such case, the energies of bound anyon pairs scaling as $2U$ are well-separated from the rest of the two-particle states which effectively reduces the dimensionality of the problem yielding a one-dimensional system with the alternating effective coupling amplitudes $J_1^{(\rm{eff})}=|J^2\,e^{i\theta}/U+P|$ and $J_2^{(\rm{eff})}=J^2/|U|$. Here, the effective coupling constant $J_1^{(\rm{eff})}$ is determined by the interference of the single-particle tunneling and direct two-particle hopping, while the statistical exchange angle provides the phase shift between the two contributions. Thus, the ratio of the effective coupling constants can be larger or smaller than 1 depending on the value of $\theta$ which according to the well-known physics of the Su-Schrieffer-Heeger model~\cite{1979_SSH} indicates the topological transitions accompanied by the topological edge states.

To confirm this prediction, we investigate the energy spectrum and two-particle wave functions in our model for the different values of statistical exchange angle $\theta$ by exact diagonalization of Eqs.~\eqref{eq:Tight_binding-1}-\eqref{eq:Tight_binding-3}. The energy diagram calculated for $\theta \in [0,\pi]$ is shown in Fig.~\ref{fig:Theory}(a). To quantify the localization of the eigenstates, we calculate their inverse participation ratios (IPR)~\cite{1974_Thouless}
\begin{equation}
    {\rm IPR} = \sum_{n,m}|\beta_{mn}|^4,
    \label{eq:IPR}
\end{equation}
where the summation is performed over all sites of the system and the wave function is normalized to unity. In such case, ${\rm IPR} \to 1$ corresponds to the strongly localized excitations, while the opposite scenario ${\rm IPR} \to 0$ describes  delocalized eigenstates.

Eigenmodes calculated for the different types of quantum statistics feature several types of the two-particle states. The majority of the states form a continuum of strongly delocalized {\it scattering states}, Fig.~\ref{fig:Theory}(d,g,j). In the case of a periodic array, their energy is given by the sum of the respective single-particle energies being unaffected by the interactions. 

At higher energies, the scattering continuum is followed by the two bands of bound particle pairs, {\it doublons}~\cite{2006_Winkler}, featuring co-localization of anyons in the same or adjacent sites, as it is evident from the associated probability distributions~Fig.~\ref{fig:Theory}(b,e,h). Such co-localization is also manifested in the values of the inverse participation ratio distinct from those of the scattering states [Fig.~\ref{fig:Theory}(a)]. While the relative positions of the particles are restricted, the pair as a whole can move freely along the array.

Finally, there is a single mode located inside the doublon bandgap with the probability distributions depicted in Fig.~\ref{fig:Theory}(c,i). In such state, the anyon pair is pinned to a single site of the lattice forming doublon edge state. The very existence of this state is linked to the topology of the doublon bands as evidenced by the direct calculation of the topological invariant (Supplementary Materials, Sec.~II).

Most remarkably, doublon bandstructure evolves with the statistical angle $\theta$ featuring closing and reopening of a bandgap in the vicinity of $\theta\approx 1$ point accompanied by the change of the topological invariant. At the same time, doublon edge state shifts from the leftmost corner of the array [Fig.~\ref{fig:Theory}(c)] to the rightmost one [Fig.~\ref{fig:Theory}(i)] delocalizing in the topological transition point [Fig.~\ref{fig:Theory}(f)]. This behavior is a fingerprint of topological transition driven by the quantum statistics $\theta$ of interacting anyons.

%_________________________Figure_3__________________________
\begin{figure*}[ht]
    \centering
    \includegraphics[width=15cm]{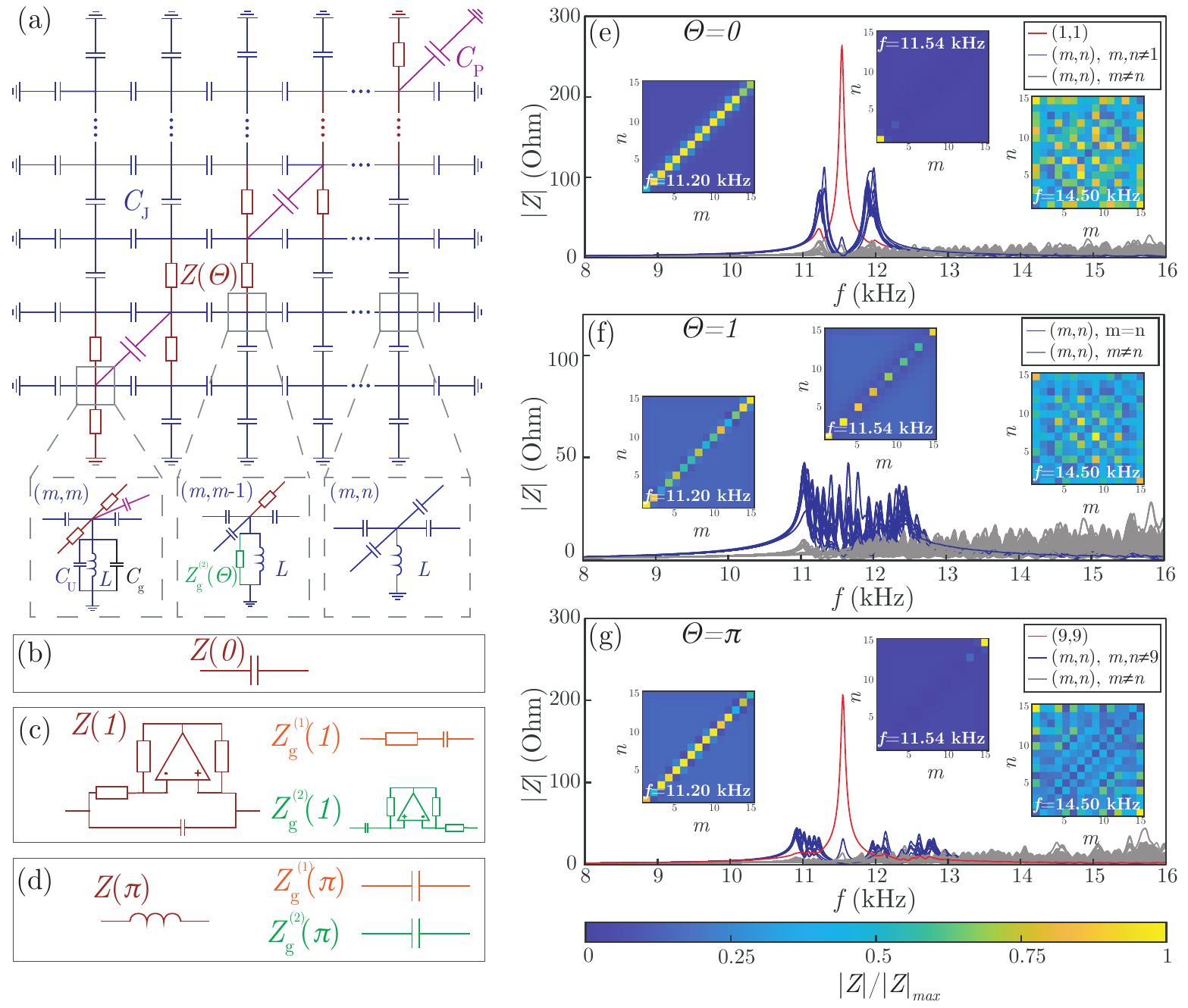}
    \caption{Analog emulator of the proposed system based on resonant electric circuits. Equivalent electric circuit (a) consists of capacitors $C_{\rm J}=1$~$\mu {\rm F}$, $C_{\rm P}=0.75$~$\mu {\rm F}$, $C_{\rm U}=2.25$~$\mu {\rm F}$, inductors $L=23.21$~$\mu {\rm H}$, and impedances $Z(\theta)$, $Z^{(1)}_{g}(\theta)$ which depend on the quantum statistics. (b) For bosonic statistics $\theta=0$, the element $Z$ is a capacitor $C_J$, and grounding elements $Z^{(1)}_{g}$ and $Z^{(2)}_{g}$ are not required. (c) For fractional quantum statistics $\theta =1$, $Z$ is a parallel connection of capacitance $C=0.48\,\mu$F and negative impedance converter (NIC) with $R=\pm 15.8$ Ohm, where the sign depends on direction, while $Z^{(1)}_g$ and $Z^{(2)}_g$ are a serial connection of a resistor $R=11.77$ Ohm and a capacitor $C=2 \mu F$ or a negative resistor $R=-11.77$ Ohm (implemented with NIC) and a capacitor $C=2\,\mu$F, respectively. Positive and negative resistance in the circuit is balanced. (d) For the fermionic statistics $\theta = \pi$, $Z$ can be approximated as inductor $L_J=190.21\,\mu$H, while $Z^{(1)}_g$ and $Z^{(2)}_g$ are $2\,\mu$F capacitors. (e-g) The results of numerical simulations of the circuits with $15 \times 15$ nodes for the cases $\theta=0$, $\theta=1$ and $\theta=\pi$, respectively. Different curves correspond to the impedance spectra of the different circuit nodes $(m,n)$, insets show the impedance distribution at specific frequencies.}  
    \label{fig:Circuit}
\end{figure*}

%_________________________Figure_3__________________________

%_______________________Electric_circuit_______________________
{\it Electric circuit emulation.~}---~As the direct experimental verification of the predicted statistics-induced topological transitions is challenging, we design an analog emulator of our quantum system based on resonant electric circuits. Within such an approach, tight-binding equations \eqref{eq:Tight_binding-1}-\eqref{eq:Tight_binding-3} are rigorously mapped onto the set of Kirchhoff's rules for every node $i$ of the circuit $\sum_{j}I_{ij} = \sum_{j}\sigma_{ij}(\varphi_i - \varphi_j) = 0$, where the indices $i,j=1...N^{2}$ enumerate the nodes of the circuit, $I_{ij}$ is the current flowing into $i$-th node from $j$-th node, $\sigma_{ij}$ is the admittance between the respective nodes, and $\varphi_{j}$ is the electric potential at the node $j$. 

The desired mapping is realized by the circuit shown in Fig.~\ref{fig:Circuit}(a) while the parameters of the model are defined in terms of the circuit elements as follows (Supplementary Materials, Sec.~III):
\begin{equation}
   \frac{U}{J}=\dfrac{C_{\rm P}+C_{\rm U}}{2C_{\rm J}}, \quad \frac{P}{J}=-\dfrac{C_{\rm P}}{C_{\rm J}}, \quad \frac{\varepsilon}{J} = \frac{f_{0}^2}{f^2} - 4,
\end{equation}
where $C_{\rm J}$, $C_{\rm U}$, and $C_{\rm P}$ are the capacitances constituting the circuit, $f_{0}=1/(2\,\pi\sqrt{LC_{\rm J}})$ and $L$ is the inductance of grounding inductors placed at every node of the circuit. Thus, low energies $\varepsilon$ in the tight-binding model correspond to high frequencies $f$, and vice versa. Note also that some additional grounding elements should be introduced at the edges and corners to maintain the correspondence between the tight-binding model and circuit equations (Supplementary Materials, Sec.~III).

While real coupling links $J$ and $P$ are readily realized with the help of capacitors $C_{\rm J}$ and $C_{\rm P}$, implementation of complex coupling links $J\,e^{\pm i\theta}$ requires specially designed impedances $Z(\theta)$ including  in the general case such active components as operational amplifiers [Fig.~\ref{fig:Circuit}(c)]. The designed setup simplifies in the limiting cases $\theta=0$ and $\theta=\pi$ when the impedance $Z(\theta)$ corresponds either to positive [Fig.~\ref{fig:Circuit}(b)] or negative capacitance [Fig.~\ref{fig:Circuit}(d)], the latter can be approximated by the suitable inductors.

Using the designs of $Z(\theta)$ elements shown in Fig.~\ref{fig:Circuit}(b-d), we simulate the three circuits corresponding to the representative cases $\theta=0$ (bosons), $\theta=1$ (anyons, topological transition point) and $\theta=\pi$ (pseudo-fermions). In our simulations performed with Keysight Advanced Design System we retrieve the frequency dependence of the circuit impedance when the voltage is applied to the different sites $(m,n)$ [Fig.~\ref{fig:Circuit}(e-g)]. In such case, characteristic peaks in the impedance spectra correspond to the circuit modes with the nonzero amplitude in a given site. To identify the type of the mode behind a specific peak, we calculate the two-dimensional map of circuit impedance at a given frequency for the various positions $(m,n)$ of the voltage source. Comparing this map with the theoretical probability distributions $|\beta_{mn}|^2$, we discriminate the modes that correspond to the scattering states, doublons as well as doublon edge states.

In particular, our calculations [Fig.~\ref{fig:Circuit}(e-g) reveal the bandgap around $11.5$~kHz frequency present for bosonic and fermionic cases but absent in the topological transition point $\theta=1$. The closest bulk bands feature impedance maxima for the diagonal sites $(n,n)$ indicating doublon modes, while higher-frequency peaks correspond to the lower-energy scattering states. Moreover, both circuits with $\theta=0$ and $\theta=\pi$ harbor a single resonant peak inside the bandgap [Fig.~\ref{fig:Circuit}(e,g)]. The associated impedance map has a sharp maximum at a single pixel, which has different positions in bosonic and fermionic cases. These results confirm our two crucial predictions: closing and reopening of a doublon bandgap as well as change a in the localization of the edge mode.

%__________________________Conclusion__________________________
To conclude, we have predicted a topological transition in the two-particle system driven by the quantum statistics of interacting anyons. While anyonic statistics is challenging to manipulate in real experiments, we put forward an alternative avenue based on analog emulators realized as resonant electric circuits. Our results thus provide interesting perspective on topological transitions in many-body interacting systems.

\begin{acknowledgments}
We thank Alexander Poddubny and Pavel Seregin for valuable discussions. This work was supported by the Russian Science Foundation (grant No. 21-72-10107). N.O. acknowledges partial support by the Foundation for the Advancement of Theoretical Physics and Mathematics ``BASIS''.
\end{acknowledgments}

%__________________________References___________________________

%\bibliography{Anyons_BibFile}

%

\end{document}

%\preprint{APS/123-QED}

\title{Supplementary Information\\~\\Electric circuit emulation of topological transitions driven by quantum statistics}

\author{Nikita A. Olekhno}
\thanks{These two authors contributed equally}
%\email{nikita.olekhno@metalab.ifmo.ru}
\affiliation{School of Physics and Engineering, ITMO University, Saint Petersburg 197101, Russia}
\author{Alina D. Rozenblit}
\thanks{These two authors contributed equally}
\affiliation{School of Physics and Engineering, ITMO University, Saint Petersburg 197101, Russia}
\thanks{These two authors contributed equally}
\author{Alexey A. Dmitriev}
\affiliation{School of Physics and Engineering, ITMO University, Saint Petersburg 197101, Russia}
\author{Daniel A. Bobylev}
\affiliation{School of Physics and Engineering, ITMO University, Saint Petersburg 197101, Russia}
\author{Maxim A. Gorlach}
% \email{m.gorlach@metalab.ifmo.ru}
\affiliation{School of Physics and Engineering, ITMO University, Saint Petersburg 197101, Russia}

%\date{\today}

\maketitle

\tableofcontents

%____________________________The_model______________________________
\section{Theoretical model}

In this Section, we consider the basic properties of the one-dimensional (1D) extended two-particle Hubbard Hamiltonian
\begin{multline}
    \hat{H}= \omega_{0}\sum_{m=1}^{N}\hat{n}_{m} + U\,\sum_{m=1}^{N} \hat{n}_{m}(\hat{n}_{m}-1) - J\sum_{m=1}^{N-1}(\hat{a}^{\dag}_{m}\hat{a}_{m+1} + \hat{a}^{\dag}_{m+1}\hat{a}_{m}) + \\ + \dfrac{P}{2}\sum_{j=1}^{(N-1)/2}(\hat{a}^{\dag}_{2j-1}\hat{a}^{\dag}_{2j-1}\hat{a}_{2j}\hat{a}_{2j} + \hat{a}^{\dag}_{2j}\hat{a}^{\dag}_{2j}\hat{a}_{2j-1}\hat{a}_{2j-1}),
\label{eq:Hamiltonian_Supp}
\end{multline}
in the case of fractional quantum statistics for creation and annihilation operators $\hat{a}^{\dag}$, $\hat{a}$:
\begin{align}
  \hat{a}_{l}\hat{a}_{k}&=\exp(i\theta\, {\rm sgn}(l-k))\,\hat{a}_{k}\hat{a}_{l},\\
   \hat{a}_{l}\hat{a}^{\dag}_{k}&=\delta_{lk} + \exp(-i \theta\, {\rm sgn}(l-k))\,\hat{a}^{\dag}_{k}\hat{a}_{l},
  \label{eq:Commutators_Supp}
\end{align}
where $\theta$ is the statistical exchange angle.

%_____________Particle_number_conservation_____________
\subsection{Particle number conservation}
\label{sec:Particle_number}

First, we demonstrate that the Hamiltonian Eq.(\ref{eq:Hamiltonian_Supp}) commutes with the particle number operator $\hat{n}=\sum_{m}\hat{a}_{m}^{\dag}\hat{a}_{m}$, so that the total number of particles in the system is conserved. For convenience, we present the Hamiltonian Eq.(\ref{eq:Hamiltonian_Supp}) in the form
\begin{equation}
    \hat{H} = \hat{H}_{0} + \hat{H}_{U} + \hat{H}_{J} + \hat{H}_{P},
\end{equation}
where
\begin{align}
    \hat{H}_{0} &= \omega_{0}\sum_{m=1}^{N}\hat{a}_{m}^{\dag}\hat{a}_{m},\\
    \hat{H}_{U} &= U\sum_{m=1}^{N} \hat{n}_{m}(\hat{n}_{m}-1),\\
    \hat{H}_{J} &= - J\sum_{m=1}^{N-1} (\hat{a}^{\dag}_{m}\hat{a}_{m+1} + \hat{a}^{\dag}_{m+1}\hat{a}_{m}),\\
    \hat{H}_{P} &= \dfrac{P}{2}\sum_{j=1}^{(N-1)/2}(\hat{a}^{\dag}_{2j-1}\hat{a}^{\dag}_{2j-1}\hat{a}_{2j}\hat{a}_{2j} + \hat{a}^{\dag}_{2j}\hat{a}^{\dag}_{2j}\hat{a}_{2j-1}\hat{a}_{2j-1}).
\end{align}
Obviously,
\begin{equation}
    [\hat{H}_{0},\hat{n}] = \omega_{0}[\hat{n},\hat{n}] = 0
\end{equation}
independently of quantum statistics. The remaining commutators can be evaluated by successively applying the relation
\begin{equation}
    [\hat{A}\hat{B},\hat{C}] = \hat{A}[\hat{B},\hat{C}] + [\hat{A},\hat{C}]\hat{B}
\end{equation}
which holds for arbitrary operators $\hat{A}$, $\hat{B}$, and $\hat{C}$. To do so, we evaluate several auxiliary commutators. As follows from the definition Eq.(\ref{eq:Commutators_Supp}),
\begin{align}
    [\hat{a}_{l},\hat{a}_{k}] &= ({\rm exp}(i\theta_{\rm sgn}(l-k)) - 1)\hat{a}_{k}\hat{a}_{l},\\
    [\hat{a}_{l}^{\dag},\hat{a}_{k}] &= -\delta_{l,k} - ({\rm exp}(i\theta_{\rm sgn}(l-k)) - 1)\hat{a}_{l}^{\dag}\hat{a}_{k}.
\end{align}
Then,
\begin{equation*}
    [\hat{n}_{l},\hat{a}_{k}] = \hat{a}_{l}^{\dag}[\hat{a}_{l},\hat{a}_{k}] + [\hat{a}_{l}^{\dag},\hat{a}_{k}]\hat{a}_{l} = -\delta_{l,k}\hat{a}_{l},
\end{equation*}
leading to the following set of commutators for creation/annihilation operators at site $l$ and particle number operator $\hat{n}_{k}$ at site $k$:
\begin{align}
    [\hat{a}_{l},\hat{n}_{k}] &= \delta_{l,k}\hat{a}_{k},\\
    [\hat{a}_{l}^{\dag},\hat{n}_{k}] &= -\delta_{l,k}\hat{a}_{k}^{\dag}.
    \label{eq:Commutator_an}
\end{align}
The next order of commutators is represented by the commutator of two particle number operators $\hat{n}_{k}$,$\hat{n}_{l}$ at sites $k$ and $l$, respectively:
\begin{equation}
    [\hat{n}_{k},\hat{n}_{l}] = [\hat{a}_{k}^{\dag}\hat{a}_{k},\hat{n}_{l}] = \hat{a}_{k}^{\dag}[\hat{a}_{k},\hat{n}_{l}] + [\hat{a}_{k}^{\dag},\hat{n}_{l}]\hat{a}_{k} = \delta_{l,k}\hat{a}_{k}^{\dag}\hat{a}_{l} - \delta_{l,k}\hat{a}_{l}^{\dag}\hat{a}_{k} = 0.
    \label{eq:Commutator_nn}
\end{equation}
From Eq.(\ref{eq:Commutator_nn}) we obtain
\begin{equation}
    [\hat{H}_{U},\hat{n}] = U\left[\sum_{m=1}^{N}\hat{n}_{m}(\hat{n}_{m} - 1), \sum_{l=1}^{N}\hat{n}_{l}\right] = U\sum_{m,l=1}^{N}\big([\hat{n}_{m}\hat{n}_{m},\hat{n}_{l}] - [\hat{n}_{m},\hat{n}_{l}]\big) = 0.
\end{equation}

For the next commutator $[\hat{H}_{J},\hat{n}]$ we obtain
\begin{multline}
    [\hat{H}_{J},\hat{n}] = \sum_{m,l=1}^{N}[\hat{a}_{m}^{\dag}\hat{a}_{m+1} + \hat{a}_{m+1}^{\dag}\hat{a}_{m},\hat{n}_{l}] = \sum_{m,l=1}^{N}\big(\hat{a}_{m}^{\dag}[\hat{a}_{m+1},\hat{n}_{l}] + [\hat{a}_{m}^{\dag},\hat{n}_{l}]\hat{a}_{m+1} + \hat{a}_{m+1}^{\dag}[\hat{a}_{m},\hat{n}_{l}] + \\ + [\hat{a}_{m+1}^{\dag},\hat{n}_{l}]\hat{a}_{m}\big) = \sum_{m,l=1}^{N}\big(\delta_{m+1,l}\hat{a}_{m}^{\dag}\hat{a}_{l} - \delta_{m,l}\hat{a}_{l}^{\dag}\hat{a}_{m+1} + \delta_{m,l}\hat{a}_{m+1}^{\dag}\hat{a}_{l} - \delta_{m+1,l}\hat{a}_{l}^{\dag}\hat{a}_{m}\big) = 0
    \label{eq:Commutator_HJ}
\end{multline}

To evaluate the last commutator $[\hat{H}_{P},\hat{n}]$, we start with the following auxiliary ones:
\begin{align}
    &[\hat{a}_{2j}\hat{a}_{2j}, \hat{n}_{l}] = \hat{a}_{2j}[\hat{a}_{2j},\hat{n}_{l}] + [\hat{a}_{2j},\hat{n}_{l}]\hat{a}_{2j} = \delta_{2j,l}(\hat{a}_{2j}\hat{a}_{l} + \hat{a}_{l}\hat{a}_{2j}),\\
    &[\hat{a}_{2j-1}\hat{a}_{2j-1}, \hat{n}_{l}] = \delta_{2j,l}(\hat{a}_{2j-1}\hat{a}_{l} + \hat{a}_{l}\hat{a}_{2j-1}). 
\end{align}
Then,
\begin{align}
    &[\hat{a}_{2j-1}^{\dag}\hat{a}_{2j}\hat{a}_{2j}, \hat{n}_{l}] = \hat{a}_{2j-1}^{\dag}[\hat{a}_{2j}\hat{a}_{2j}, \hat{n}_{l}] + [\hat{a}_{2j-1}^{\dag}, \hat{n}_{l}]\hat{a}_{2j}\hat{a}_{2j} = \delta_{2j,l}\hat{a}_{2j-1}^{\dag}(\hat{a}_{2j}\hat{a}_{l} + \hat{a}_{l}\hat{a}_{2j}) - \delta_{2j-1,l}\hat{a}_{l}^{\dag}\hat{a}_{2m}\hat{a}_{2m},\\
    &[\hat{a}_{2j}^{\dag}\hat{a}_{2j-1}\hat{a}_{2j-1}, \hat{n}_{l}] = \delta_{2j-1,l}\hat{a}_{2j}^{\dag}(\hat{a}_{2j-1}\hat{a}_{l} + \hat{a}_{l}\hat{a}_{2j-1}) - \delta_{2j,l}\hat{a}_{l}^{\dag}\hat{a}_{2m-1}\hat{a}_{2m-1},
\end{align}
and
\begin{align}
    [\hat{a}_{2j-1}^{\dag}\hat{a}_{2j-1}^{\dag}\hat{a}_{2j}\hat{a}_{2j}, \hat{n}_{l}] &= \hat{a}_{2j-1}^{\dag}[\hat{a}_{2j-1}^{\dag}\hat{a}_{2j}\hat{a}_{2j}, \hat{n}_{l}] + [\hat{a}_{2j-1}^{\dag},\hat{n}_{l}]\hat{a}_{2j-1}^{\dag}\hat{a}_{2j}\hat{a}_{2j} =\\
    &= \delta_{2j,l}\hat{a}_{2j-1}^{\dag}\hat{a}_{2j-1}^{\dag}(\hat{a}_{2j}\hat{a}_{l} + \hat{a}_{l}\hat{a}_{2j}) - \delta_{2j-1,l}(\hat{a}_{2j-1}^{\dag}\hat{a}_{l}^{\dag} + \hat{a}_{l}^{\dag}\hat{a}_{2j-1}^{\dag})\hat{a}_{2j}\hat{a}_{2j},\\
    [\hat{a}_{2j}^{\dag}\hat{a}_{2j}^{\dag}\hat{a}_{2j-1}\hat{a}_{2j-1}, \hat{n}_{l}] &= \delta_{2j-1,l}\hat{a}_{2j}^{\dag}\hat{a}_{2j}^{\dag}(\hat{a}_{2j-1}\hat{a}_{l} + \hat{a}_{l}\hat{a}_{2j-1}) - \delta_{2j,l}(\hat{a}_{2j}^{\dag}\hat{a}_{l}^{\dag} + \hat{a}_{l}^{\dag}\hat{a}_{2j}^{\dag})\hat{a}_{2j-1}\hat{a}_{2j-1}.
\end{align}
The above relations yield
\begin{multline}
    [\hat{H}_{P},\hat{n}] = \frac{P}{2} \sum_{j=1}^{(N-1)/2}\sum_{l=1}^{N}[(\hat{a}_{2j-1}^{\dag}\hat{a}_{2j-1}^{\dag}\hat{a}_{2j}\hat{a}_{2j} + \hat{a}_{2j}^{\dag}\hat{a}_{2j}^{\dag}\hat{a}_{2j-1}\hat{a}_{2j-1}),\hat{n}_{l}] =\\
    =\frac{P}{2} \sum_{j=1}^{(N-1)/2}\big(\hat{a}_{2j-1}^{\dag}\hat{a}_{2j-1}^{\dag}(\hat{a}_{2j}\hat{a}_{2j} + \hat{a}_{2j}\hat{a}_{2j}) - \hat{a}_{2j-1}^{\dag}\hat{a}_{2j-1}^{\dag}\hat{a}_{2j}\hat{a}_{2j} - \hat{a}_{2j-1}^{\dag}\hat{a}_{2j-1}^{\dag}\hat{a}_{2j}\hat{a}_{2j} +\\
    +\hat{a}_{2j}^{\dag}\hat{a}_{2j}^{\dag}(\hat{a}_{2j-1}\hat{a}_{2j-1} + \hat{a}_{2j-1}\hat{a}_{2j-1}) - \hat{a}_{2j}^{\dag}\hat{a}_{2j}^{\dag}\hat{a}_{2j-1}\hat{a}_{2j-1} - \hat{a}_{2j}^{\dag}\hat{a}_{2j}^{\dag}\hat{a}_{2j-1}\hat{a}_{2j-1}\big) = 0.
\end{multline}
Finally, combining all the above partial commutators for operators $\hat{H}_{0}$, $\hat{H}_{U}$, $\hat{H}_{J}$, and $\hat{H}_{P}$, we obtain
\begin{equation}
    [\hat{H},\hat{n}] = 0.
\end{equation}

The conservation of quasi-particle number allows us to search the wave function in the form
\begin{equation}
    \ket{\psi} = \frac{1}{\sqrt{2}} \sum_{m,n=1}^{N}\beta_{mn}\hat{a}_{m}^{\dag}\hat{a}_{n}^{\dag}\ket{0}
    \label{eq:psi_Supp}
\end{equation}
used in the main text.

%_______________Wavefunction_symmetry_____________
\subsection{Wave function symmetry}
\label{sec:Symmetry}

Next, we characterize the symmetry properties of superposition coefficients $\beta_{mn}$ in the case of fractional statistics defined by Eq.~(\ref{eq:Commutators_Supp}). As follows from the form of wave function,  Eq.~(\ref{eq:psi_Supp}),
\begin{equation*}
    \ket{\psi} = \frac{1}{\sqrt{2}} \sum_{m,n=1}^{N}\beta_{mn}\hat{a}_{m}^{\dag}\hat{a}_{n}^{\dag}\ket{0} = \frac{1}{\sqrt{2}} \sum_{m,n=1}^{N}\beta_{mn}{\rm exp}\big(i\theta_{\rm sgn}(m-n)\big)\hat{a}_{n}^{\dag}\hat{a}_{m}^{\dag}\ket{0} \equiv \frac{1}{\sqrt{2}} \sum_{m,n=1}^{N}\beta_{nm}\hat{a}_{n}^{\dag}\hat{a}_{m}^{\dag}\ket{0},
\end{equation*}
leading to
\begin{equation}
    \beta_{nm} = \beta_{mn}{\rm exp}\big(i\theta_{\rm sgn}(m-n)\big).
    \label{eq:beta_symm_Supp}
\end{equation}

%_______________Tight_binding_equations_____________
\subsection{Tight-binding equations}
\label{sec:Tight_binding}

Given the Hamiltonian Eq.~(\ref{eq:Hamiltonian_Supp}) and the wave function Eq.~(\ref{eq:psi_Supp}), we write down the tight-binding equations for the superposition coefficients $\beta_{mn}$ that follow from the Schr\"odinger equation  $\hat{H}\ket{\psi}=E\ket{\psi}$. To proceed, we present the Hamiltonian as a sum of  $\hat{H}_{0}$, $\hat{H}_{U}$, $\hat{H}_{J}$, and $\hat{H}_{P}$ contributions. For the first one we obtain
\begin{multline}
    \hat{H}_{0}\ket{\psi} = \omega_{0}\frac{1}{\sqrt{2}}\sum_{m,n=1}^{N}\sum_{l=1}^{N}\beta_{mn}\hat{n}_{l}\hat{a}_{m}^{\dag}\hat{a}_{n}^{\dag}\ket{0} = \omega_{0}\frac{1}{\sqrt{2}}\sum_{m,n,l=1}^{N}\beta_{mn}\hat{a}_{l}^{\dag}\hat{a}_{l}\hat{a}_{m}^{\dag}\hat{a}_{n}^{\dag}\ket{0} =\\ = \omega_{0}\frac{1}{\sqrt{2}}\sum_{m,n,l=1}^{N}\beta_{mn}\hat{a}_{l}^{\dag}\big(\delta_{l,m} + \exp{-i\theta_{\rm sgn}(l-m)}\hat{a}_{m}^{\dag}\hat{a}_{l}\big)\hat{a}_{n}^{\dag}\ket{0} = \omega_{0}\frac{1}{\sqrt{2}}\sum_{m,n=1}^{N}\beta_{mn}\hat{a}_{m}^{\dag}\hat{a}_{n}^{\dag}\ket{0} +\\
    + \omega_{0}\frac{1}{\sqrt{2}}\sum_{m,n,l=1}^{N}\beta_{mn}\exp(-i\theta_{\rm sgn}(l-m))\hat{a}_{l}^{\dag}\hat{a}_{m}^{\dag}\big(\delta_{l,n} + \exp(-i\theta_{\rm sgn}(l-n))\hat{a}_{n}^{\dag}\hat{a}_{l}\big)\ket{0} = \omega_{0}\frac{1}{\sqrt{2}}\sum_{m,n=1}^{N}\beta_{mn}\hat{a}_{m}^{\dag}\hat{a}_{n}^{\dag}\ket{0} +\\
    + \omega_{0}\frac{1}{\sqrt{2}}\sum_{m,n=1}^{N}\beta_{mn}\exp(i\theta_{\rm sgn}(m-n))\hat{a}_{n}^{\dag}\hat{a}_{m}^{\dag}\ket{0} = 2\omega_{0}\frac{1}{\sqrt{2}}\sum_{m,n=1}^{N}\beta_{mn}\hat{a}_{m}^{\dag}\hat{a}_{n}^{\dag}\ket{0},
    \label{eq:TB_H_0}
\end{multline}
which results in a trivial shift of the eigenenergies $E$ by the constant value of $2\omega_{0}$. In the following, we consider shifted eigenenergies $\varepsilon = E - 2\omega_{0}$.

To proceed with the second term $\hat{H}_{U}\ket{\psi}$, we apply the commutation relation
\begin{equation}
    [\hat{a}_{m}^{\dag}\hat{a}_{k}^{\dag},\hat{n}_{l}] = \hat{a}_{m}^{\dag}[\hat{a}_{k}^{\dag},\hat{n}_{l}] + [\hat{a}_{m}^{\dag},\hat{n}_{l}]\hat{a}_{k}^{\dag} = -\delta_{l,k}\hat{a}_{m}^{\dag}\hat{a}_{l}^{\dag} - \delta_{m,l}\hat{a}_{l}^{\dag}\hat{a}_{k}^{\dag},
\end{equation}
and evaluate the following operators:
\begin{align}
    &\hat{n}_{l}\hat{a}_{m}^{\dag}\hat{a}_{k}^{\dag}\ket{0} = \big(\hat{a}_{m}^{\dag}\hat{a}_{k}^{\dag}\hat{n}_{l} + \delta_{l,k}\hat{a}_{m}^{\dag}\hat{a}_{l}^{\dag} + \delta_{m,l}\hat{a}_{l}^{\dag}\hat{a}_{k}^{\dag}\big)\ket{0} = \big(\delta_{l,k}\hat{a}_{m}^{\dag}\hat{a}_{l}^{\dag} + \delta_{m,l}\hat{a}_{l}^{\dag}\hat{a}_{k}^{\dag}\big)\ket{0},\\
    &\hat{n}_{l}\hat{n}_{l}\hat{a}_{m}^{\dag}\hat{a}_{k}^{\dag}\ket{0} = \big(\delta_{n,l}\hat{a}_{m}^{\dag}\hat{a}_{l}^{\dag} + \delta_{n,l}\delta_{m,l}\hat{a}_{l}^{\dag}\hat{a}_{l}^{\dag} + \delta_{n,l}\delta_{m,l}\hat{a}_{m}^{\dag}\hat{a}_{l}^{\dag} + \delta_{m,l}\hat{a}_{l}^{\dag}\hat{a}_{n}^{\dag}\big)\ket{0}.
\end{align}
Then,
\begin{multline}
    \hat{H}_{U}\ket{0} = U\frac{1}{\sqrt{2}}\sum_{m,n,l=1}^{N}\beta_{mn}\hat{n}_{l}(\hat{n}_{l} - 1)\hat{a}_{m}^{\dag}\hat{a}_{n}^{\dag}\ket{0} = U\frac{1}{\sqrt{2}}\sum_{m,n,l=1}^{N}\beta_{mn}\big(\delta_{n,l}\hat{a}_{m}^{\dag}\hat{a}_{l}^{\dag} + \delta_{n,l}\delta_{m,l}\hat{a}_{l}^{\dag}\hat{a}_{l}^{\dag} + \delta_{n,l}\delta_{m,l}\hat{a}_{m}^{\dag}\hat{a}_{l}^{\dag} +\\
    + \delta_{m,l}\hat{a}_{l}^{\dag}\hat{a}_{n}^{\dag} - \delta_{n,l}\hat{a}_{m}^{\dag}\hat{a}_{l}^{\dag} - \delta_{m,l}\hat{a}_{l}^{\dag}\hat{a}_{n}^{\dag}\big)\ket{0} = U\frac{1}{\sqrt{2}}\sum_{m,n=1}^{N}\beta_{mn}\big(\hat{a}_{m}^{\dag}\hat{a}_{n}^{\dag} + \delta_{m,n}\hat{a}_{m}^{\dag}\hat{a}_{n}^{\dag} + \delta_{m,n}\hat{a}_{m}^{\dag}\hat{a}_{n}^{\dag} + \hat{a}_{m}^{\dag}\hat{a}_{n}^{\dag} -\\
    -\hat{a}_{m}^{\dag}\hat{a}_{n}^{\dag} - \hat{a}_{m}^{\dag}\hat{a}_{n}^{\dag}\big)\ket{0} = 2U\frac{1}{\sqrt{2}}\sum_{m,n=1}^{N}\beta_{mn}\delta_{m,n}\hat{a}_{m}^{\dag}\hat{a}_{n}^{\dag}\ket{0}.
    \label{eq:TB_H_U}
\end{multline}

For the third term $\hat{H}_{J}\ket{\psi}$ we start with the following operators:
\begin{align}
    &\hat{a}_{l}^{\dag}\hat{a}_{l+1}\hat{a}_{m}^{\dag}\hat{a}_{n}^{\dag}\ket{0} = \big(\delta_{l+1,m}\hat{a}_{l}^{\dag}\hat{a}_{n}^{\dag} + \delta_{l+1,n}\exp(-i\theta_{\rm sgn}(l+1-m))\hat{a}_{l}^{\dag}\hat{a}_{m}^{\dag}\big)\ket{0},\\
    &\hat{a}_{l+1}^{\dag}\hat{a}_{l}\hat{a}_{m}^{\dag}\hat{a}_{n}^{\dag}\ket{0} = \big(\delta_{l,m}\hat{a}_{l+1}^{\dag}\hat{a}_{n}^{\dag} + \delta_{l,n}\exp(-i\theta_{\rm sgn}(l-m))\hat{a}_{l+1}^{\dag}\hat{a}_{m}^{\dag}\big)\ket{0}.
\end{align}
Substituting the obtained expressions to $\hat{H}_{J}\ket{\psi}$, we obtain
\begin{multline}
    \hat{H}_{J}\ket{\psi} = -J\frac{1}{\sqrt{2}}\sum_{m,n,l=1}^{N}\beta_{mn}(\hat{a}_{l}^{\dag}\hat{a}_{l+1} + \hat{a}_{l+1}^{\dag}\hat{a}_{l})\hat{a}_{m}^{\dag}\hat{a}_{n}^{\dag}\ket{0} =\\ =-J\frac{1}{\sqrt{2}}\sum_{m,n,l=1}^{N}\beta_{mn}\big(\delta_{l+1,m}\hat{a}_{l}^{\dag}\hat{a}_{n}^{\dag} + \delta_{l+1,n}\exp(-i\theta_{\rm sgn}(l+1-m))\hat{a}_{l}^{\dag}\hat{a}_{m}^{\dag} +\\
    +\delta_{l,m}\hat{a}_{l+1}^{\dag}\hat{a}_{n}^{\dag} + \delta_{l,n}\exp(-i\theta_{\rm sgn}(l-m))\hat{a}_{l+1}^{\dag}\hat{a}_{m}^{\dag}\big)\ket{0},
\end{multline}
which, after the summation over $l$ and a shift in the indexes $m,n$ takes the following form:
\begin{multline}
    \hat{H}_{J}\ket{\psi} = -J\frac{1}{\sqrt{2}}\sum_{m,n=1}^{N}\big(\beta_{m+1,n} + \beta_{m,n+1}\exp(-i\theta({\rm sgn}(n+1-m)-{\rm sgn}(n-m))) +\\
    +\beta_{m-1,n} + \beta_{m,n-1}\exp(-i\theta({\rm sgn}(n-1-m)-{\rm sgn}(n-m)))\big)\hat{a}_{m}^{\dag}\hat{a}_{n}^{\dag}\ket{0}.
    \label{eq:TB_H_J}
\end{multline}

For the last term, the following relation is used:
\begin{equation}
    \hat{a}_{2j}\hat{a}_{2j}\hat{a}_{m}^{\dag}\hat{a}_{n}^{\dag}\ket{0} = \delta_{2j,m}\delta_{2j,n}\big(1 + \exp(-i\theta_{\rm sgn}(2j - m))\big)\ket{0}.
\end{equation}
Hence,
\begin{align}
    &\hat{a}_{2j-1}^{\dag}\hat{a}_{2j-1}^{\dag}\hat{a}_{2j}\hat{a}_{2j}\hat{a}_{m}^{\dag}\hat{a}_{n}^{\dag}\ket{0} = \delta_{2j,m}\delta_{2j,n}\big(1 + \exp(-i\theta_{\rm sgn}(2j - m))\big)\hat{a}_{2j-1}^{\dag}\hat{a}_{2j-1}^{\dag}\ket{0},\\
    &\hat{a}_{2j}^{\dag}\hat{a}_{2j}^{\dag}\hat{a}_{2j-1}\hat{a}_{2j-1}\hat{a}_{m}^{\dag}\hat{a}_{n}^{\dag}\ket{0} = \delta_{2j-1,m}\delta_{2j-1,n}\big(1 + \exp(-i\theta_{\rm sgn}(2j - 1 - m))\big)\hat{a}_{2j}^{\dag}\hat{a}_{2j}^{\dag}\ket{0},
\end{align}
and the last term reads
\begin{multline}
    \hat{H}_{P}\ket{0} = P\frac{1}{\sqrt{2}}\sum_{m,n=1}^{N}\sum_{j=1}^{(N-1)/2}\beta_{mn}(\hat{a}_{2j-1}^{\dag}\hat{a}_{2j-1}^{\dag}\hat{a}_{2j}\hat{a}_{2j} + \hat{a}_{2j}^{\dag}\hat{a}_{2j}^{\dag}\hat{a}_{2j-1}\hat{a}_{2j-1})\hat{a}_{m}^{\dag}\hat{a}_{n}^{\dag}\ket{0} =\\ P\frac{1}{\sqrt{2}}\Big(\sum_{m \in {\rm odd}}\beta_{m+1,m+1}\hat{a}_{m}^{\dag}\hat{a}_{n}^{\dag}\ket{0} + \sum_{m \in {\rm even}}\beta_{m-1,m-1}\hat{a}_{m}^{\dag}\hat{a}_{n}^{\dag}\ket{0}\Big).
    \label{eq:TB_H_P}
\end{multline}

As seen from Eqs.~(\ref{eq:TB_H_0},\ref{eq:TB_H_U},\ref{eq:TB_H_J},\ref{eq:TB_H_P}), fractional statistics manifests itself only in the term $\hat{H}_{J}\ket{\psi}$ describing the tunneling of a single anyon between two adjacent resonators. In turn, $\hat{H}_{0}$ leads to the energy shift for arbitrary $m$ and $n$, $\hat{H}_{U}$ shifts only diagonal sites with $m=n$, and two-photon tunneling also contributes to the diagonal terms only. Importantly, the statistical exchange angle $\theta$ appears in Eq.(\ref{eq:TB_H_J}) only for $m=n$ and $m=n \pm 1$, otherwise ${\rm sgn}(n+1-m)-{\rm sgn}(n-m)={\rm sgn}(n-1-m)-{\rm sgn}(n-m)=0$. Comparing the coefficients in front of $\hat{a}_{m}^{\dag}\hat{a}_{m}^{\dag}$ in the left- and right-hand sides of the eigenvalue problem $\hat{H}\ket{\psi}=E\ket{\psi}$, we finally obtain a linear system of equations
\begin{multline}
    -J\big(\beta_{m+1,n} + \beta_{m,n+1}\exp(-i\theta({\rm sgn}(n+1-m)-{\rm sgn}(n-m))) +\beta_{m-1,n} +\\ +\beta_{m,n-1}\exp(-i\theta({\rm sgn}(n-1-m)-{\rm sgn}(n-m)))\big) + P\big(\beta_{m+1,m+1}\delta_{{\rm mod}(m,2),1} +\\
    + \beta_{m-1,m-1}\delta_{{\rm mod}(m,2),0}\big)\delta_{m,n} = (E - 2\omega_{0} - 2U\delta_{m,n})\beta_{mn},
    \label{eq:TB_Supp}
\end{multline}
representing a two-dimensional (2D) tight-binding model where $(m,n)$ provide 2D coordinates of sites and $\beta_{mn}$ are the amplitudes of the 2D wave function. Note that the rigorous correspondence between 1D and 2D models is obtained only for the system eigenmodes. Taking into account the symmetry property Eq.~(\ref{eq:beta_symm_Supp}), the following sets of sites $(m,n)$  should be considered:
\begin{align*}
    &(1): \quad m=1, n=1\\
    &(2): \quad m=2, n=1\\
    &(3): \quad m=1, n=2\\
    &(4): \quad m=n, \quad 1<m,n<N, \quad m \in {\rm even}\\
    &(5): \quad m=n+1, \quad 2<m<N\\
    &(6): \quad m=n-1, \quad 1<m<N-1\\
    &(7): \quad m=n, \quad 1<m,n<N, \quad m \in {\rm odd}\\
    &(8): \quad m=N, n=N\\
    &(9): \quad 2<m<N, n=1\\
    &(10): \quad m=N, n=1\\
    &(11): \quad 1<m,n<N, \quad |m-n|>1
\end{align*}

%______________________Topological_invariant________________________
\section{Topological invariant calculation}

To examine the topological properties of our two-anyon one-dimensional system, we apply the strategy proposed in Ref.~\cite{HughesPhysRevB} and later exploited for the case of bound particle pairs in Ref.~\cite{2020_Stepanenko}. It is based on the parity analysis of the Bloch modes of the respective periodic system at time-reversal-invariant momenta $k=0$ and $k=\pi$. The Zak phase of the $n$th band of an inversion-symmetric system is calculated as~\cite{HughesPhysRevB}:
\begin{equation}
    \gamma_n = \alpha(0) - \alpha(\pi),
\end{equation}
where $\gamma_n$ is defined modulo $2\pi$ and $\alpha(k)$ describes the phase acquired by the periodic part of the wave function $\ket{u_{k}^{(n)}}$ under inversion $\hat{P}$: $\hat{P} \ket{u_{k}^{(n)}} = {\rm e}^{i\alpha(k)} \ket{u_{k}^{(n)}}$. In many cases, the analysis can be simplified to extract the Zak phase directly from the finite system simulation without the explicit derivation of the Bloch modes.

\begin{figure}[tbp]
    \centering
    \includegraphics{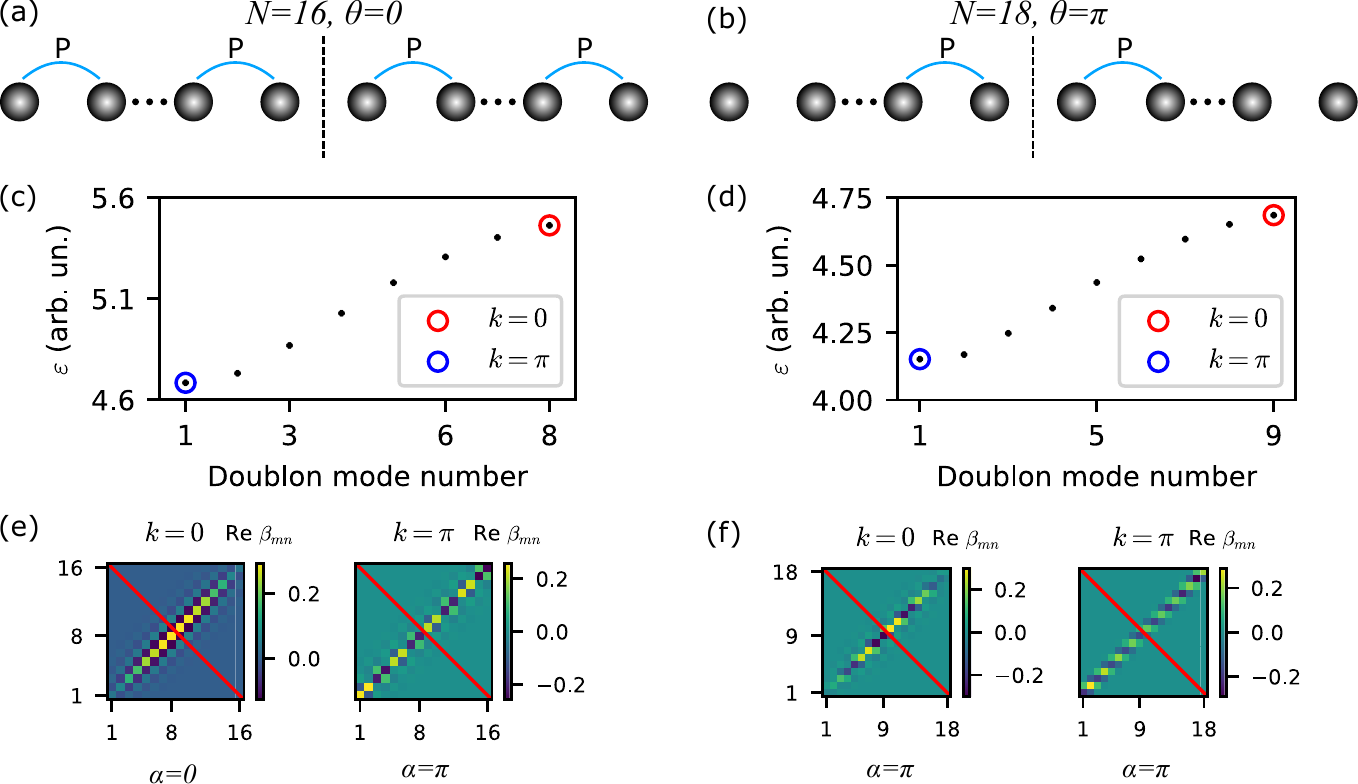}
    \caption{Topological invariant calculation. Panels (a,b) show the schemes of the structures. Array (a) having $N=16$ sites is considered for the case $\theta=0$, whereas array (b) with $N=18$ sites is taken for $\theta=\pi$. The number of sites is chosen to ensure no two-anyon hopping at the center and to avoid the emergence of edge-localized modes. Panels (c, d) depict upper-band doublon spectrum for $\theta=0, \pi$, respectively; red and blue circles denote the modes corresponding to the high-symmetry points of Brillouin zone of the respective periodic array. (e,f) demonstrate the distribution of $\mathrm{Re}\beta_{mn}$ of the circled modes for $\theta=0, \pi$.}
    \label{fig:Top_Inv}
\end{figure}

To assess the physics of the topological phase transition, we consider the modes of a finite one-dimensional array for the two values of statistical exchange angle: $\theta = 0$ and $\theta = \pi$. Since the Zak phase depends on the unit cell choice, we fix the unit cell to be without two-particle hopping $P$ in the center (for brevity, we call this type of tunneling $J$-link). The number of sites $N$ is chosen to ensure global inversion symmetry of the system. For clarity, we also choose such termination of the array that does not favour the edge modes, since these modes are not needed in the analysis of the bulk bandstructure. The simulated configurations are depicted in Fig.~\ref{fig:Top_Inv}(a,b). The rest of the tight-binding parameters correspond to Fig.~2 of the article main text: $J=1$, $U=1.5$, $P=-0.75$.

For both values of the statistical exchange angle, we calculate the modes of the finite system, selecting the upper band of doublon states [Fig.~\ref{fig:Top_Inv}(c,d)]. Next, we focus on the modes that correspond to $k=0$ and $k=\pi$ and retrieve the probability amplitudes $\beta_{mn}$ [Fig.~\ref{fig:Top_Inv}(e,f)] checking their behavior with respect to inversion which brings each site $(m,n)$ to the point $(N/2-n,N/2-m)$.

%_________________________Figure_1__________________________
%_________________________Figure_1__________________________

For the case of bosonic statistics $\theta=0$, the modes in the center and at the edge of the Brillouin zone have different parity. Hence, $\gamma = \pi - 0 = \pi$, Fig.\ref{fig:Top_Inv}(e). The Zak phase for pseudo-fermionic case $\theta=\pi$ reads $\gamma = \pi - \pi = 0$, Fig.\ref{fig:Top_Inv}(b).

These results prove that the studied two-anyon system undergoes the topological phase transition with the change of the statistical exchange angle $\theta$, while the topological edge mode moves from $J$-link edge to $J+P$ one.

%_________________________Circuit_____________________________
\section{Equations for electric circuit}

\begin{figure}[tbp]
    \centering
    \includegraphics[width=15cm]{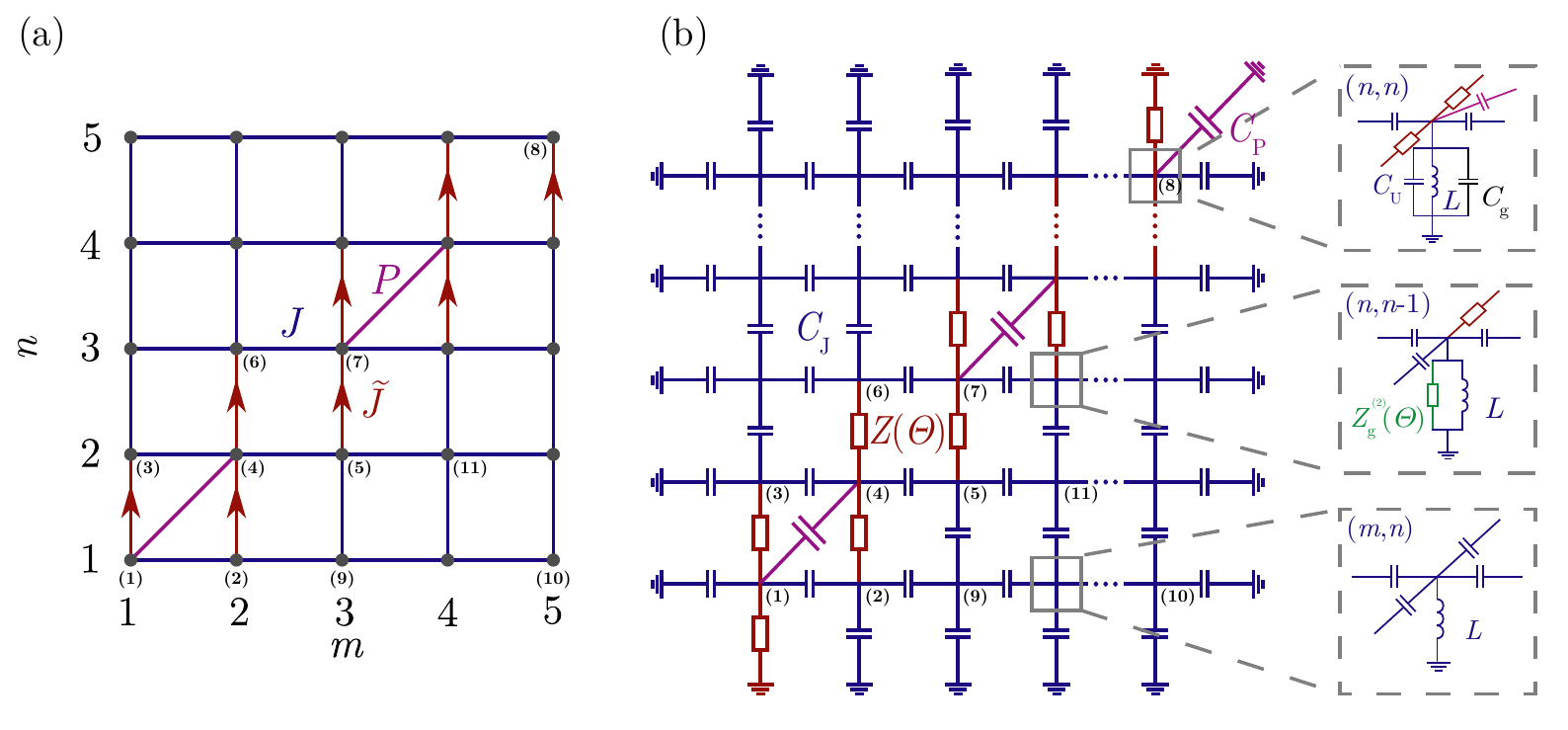}
    \caption{Mapping between the two-dimensional tight-binding model and the equivalent electrical circuit. (a) The tight-binding model includes real hopping amplitudes $J$, $P$ and complex couplings $\tilde{J}=J\,{\rm e}^{\pm i\theta}$. (b)  Resonant electrical circuit consists of capacitors $C_{\rm J}$, $C_{\rm P}$, $C_{\rm U}$, $C_{\rm g}$, inductors $L$, complex impedances $Z(\theta)$, and grounding elements $Z_{\rm g}(\theta)$ with impedance depending on the statistical exchange angle $\theta$.}
    \label{fig:Tight_binding_and_circuit}
\end{figure}

In this Section, we construct a rigorous mapping between the system of equations describing the two-dimensional tight-binding model [Fig.~\ref{fig:Tight_binding_and_circuit}(a)] and Kirchhoff's rules for currents and voltages in the equivalent circuit [Fig.~\ref{fig:Tight_binding_and_circuit}(b)].

First, we formulate the system of tight-binding equations for a set of non-identical sites indicated in Fig.~\ref{fig:Tight_binding_and_circuit}(a):
\begin{align}
    \label{eq:Mapping_TB_1}
    (1):& \quad -J{\rm e}^{-i \theta}\beta_{1,2} - J\beta_{2,1} + P\beta_{2,2} = (\varepsilon-2U_{1,1})\beta_{1,1}, \quad U_{1,1}=U+J^2/(2U)\\
    \label{eq:Mapping_TB_2}
    (2):& \quad -J(\beta_{1,1}+\beta_{3,1})-J{\rm e}^{-i \theta} \beta_{2,2}=\varepsilon\beta_{2,1}\\
    \label{eq:Mapping_TB_3}
    (3):& \quad -J(\beta_{1,3}+\beta_{2,2})-J{\rm e}^{i \theta} \beta_{1,1}=\varepsilon\beta_{1,2}\\
    \label{eq:Mapping_TB_4}
    (4):& \quad -J(\beta_{n-1,n} + \beta_{n+1,n}) - J{\rm e}^{i \theta}\beta_{n,n-1} -J{\rm e}^{-i \theta} \beta_{n,n+1} + P\beta_{n-1,n-1} = (\varepsilon-2U)\beta_{n,n}, \quad n \in {\rm even}\\
    \label{eq:Mapping_TB_5}
    (5):& \quad -J(\beta_{m+1,n} + \beta_{m-1,n} + \beta_{m,n-1}) - J{\rm e}^{-i \theta}\beta_{m,n+1} = \varepsilon \beta_{m,n}, \quad n=m-1, \quad 2<m<N\\
    \label{eq:Mapping_TB_6}
    (6):& \quad -J(\beta_{m,n+1} + \beta_{m+1,n} + \beta_{m-1,n}) - J{\rm e}^{i \theta}\beta_{m,n-1} = \varepsilon \beta_{m,n}, \quad m=n-1, \quad 2<n<N\\
    \label{eq:Mapping_TB_7}
    (7):& \quad -J(\beta_{n-1,n} + \beta_{n+1,n}) - J{\rm e}^{i \theta}\beta_{n,n-1} - J{\rm e}^{-i \theta}\beta_{n,n+1} + P\beta_{n+1,n+1} = (\varepsilon -2U)\beta_{n,n}, \quad n \in {\rm odd}\\
    \label{eq:Mapping_TB_8}
    (8):& \quad -J\beta_{N-1,N} - J{\rm e}^{i \theta}\beta_{N,N-1} = (\varepsilon - 2U_{N,N})\beta_{N,N}, \quad U_{N,N}=U+J^2/(2U)\\
    \label{eq:Mapping_TB_9}
    (9):& \quad -J(\beta_{m-1,1} + \beta_{m+1,1} +\beta_{m,2}) = \varepsilon\beta_{m,1}, \quad 2<m<N\\
    \label{eq:Mapping_TB_10}
    (10):& \quad -J(\beta_{N,2} + \beta_{N-1,1}) = \varepsilon \beta_{N,1}\\
    \label{eq:Mapping_TB_11}
    (11):& \quad -J(\beta_{m,n+1} + \beta_{m,n-1} +\beta_{m-1,n} + \beta_{m+1,n}) = \varepsilon\beta_{m,n}, \quad 1<m,n<N, \quad |m-n|>1
\end{align}
where $N$ is the size of the system, $J$ and $P$ are real tunneling couplings, $\tilde{J}$ is a complex coupling with the phase $\pm\theta$ that depends on the direction of hopping, $\varepsilon$ is an eigenstate energy, $2U$ is the energy of on-site interaction between the two particles, and $\beta_{m,n}$ is a wave function amplitude at the site $(m,n)$. Note that we have additionally introduced on-site interaction energy shifts for the sites $(1,1)$ and $(N,N)$ to prevent the formation of two-anyon Tamm-like states, which are not related to the bulk bands topology and excessively complicate the analysis. For that purpose, we choose $U_{1,1} = U_{N,N} = U+J^2/(2U)$.

%_________________________Figure_2__________________________
%_________________________Figure_2__________________________

The sites (1), (8), and (10) correspond to three different types of corners. The equations for the edge sites (2) and (3) include complex couplings $J{\rm e}^{i \theta}$ or $J{\rm e}^{-i \theta}$, while the site (9) couples with its neighbors only through the real couplings $J$. In turn, the equations describing diagonal bulk sites (4) and (7) with even and odd coordinates $(n,n)$, respectively, include all mentioned types of couplings: complex couplings with the phases $J{\rm e}^{\pm i \theta}$ as well as the real links $J$ and $P$. The convention about the sign of the coupling phase and the direction of the arrow remains the same as in the article main text. The nearest to the diagonal bulk sites (5) and (6) include only one complex coupling similarly to the edge sites (2) and (3). Finally, the bulk site (11) is surrounded solely by the real $J$ couplings. We also note that the right-hand side of the equations for the diagonal nodes $m=n$ incorporates an additional energy shift $2U$ responsible for the on-site interaction between the two particles. It is seen that the statistical exchange angle $\theta$ is present in the equations for the diagonal sites (1), (4), (7), and (8), and the nearest to the diagonal sites (2), (3), (5), and (6) as a phase parameter of complex couplings, while equations for the other sites do not include $\theta$. Hence, quantum statistics is encoded only in the diagonal region $(n,n\pm1)$ of the tight-binding model.

Next, we consider a resonant electrical circuit consisting of capacitors $C_{\rm J}$, $C_{\rm P}$, $C_{\rm U}$, $C_{\rm g}$, inductors $L$, and elements $Z$, $Z^{(1)}_{\rm g}$, and $Z^{(2)}_{\rm g}$ with impedances depending on $\theta$, [Fig.~\ref{fig:Tight_binding_and_circuit}(b)]. In such a circuit, the arrangement of elements between the nodes resembles the geometry of links in the tight-binding model. However, edge nodes of the circuit require additional groundings to compensate for the absence of neighbors and preserve the correct mapping to the tight-binding model.

According to Kirchhoff's junction rule, the sum of currents flowing into an arbitrary node $(m,n)$ vanishes: $\sum_{m',n'}I_{mn, m'n'}=0$, where $(m',n')$ are the coordinates of adjacent nodes. The above equation can be reformulated as $\sum_{m',n'}\sigma_{mn, m'n'}(\varphi_{m',n'}-\varphi_{m,n})=0$, where $\sigma_{mn, m'n'}$ is the admittance of the link between nodes $(m,n)$, and $(m'n')$, and $\varphi_{m,n}$ is the electric potential at node $(m,n)$. Such equations can be written in the following form for the nodes indicated in Fig.~\ref{fig:Tight_binding_and_circuit}(b):
\begin{align*}
    %\label{eq:Mapping_Circuit_1}
    (1):& \quad - \sigma_{\rm J}{\rm e}^{-i\theta}\varphi_{1,2} - \sigma_{\rm J}\varphi_{2,1} - \sigma_{\rm P} \varphi_{2,2} = (-2\sigma_{\rm J}(1 + \cos{\theta}) - \sigma_{\rm g} - \sigma_{\rm L} - \sigma_{\rm U} - \sigma_{\rm P})\varphi_{1,1},\\
    %\label{eq:Mapping_Circuit_2}
    (2):& \quad - \sigma_{\rm J}{\rm e}^{-i\theta}\varphi_{2,2} - \sigma_{\rm J}(\varphi_{1,1} + \varphi_{3,1}) = (-3 \sigma_{\rm J} - \sigma_{\rm J}{\rm e}^{i \theta} - \sigma_{\rm L} - \sigma_{\rm g}^{(2)})\varphi_{2,1},\\
    %\label{eq:Mapping_Circuit_3}
    (3):& \quad - \sigma_{\rm J}(\varphi_{1,3} + \varphi_{2,2}) - \sigma_{\rm J}{\rm e}^{i\theta}\varphi_{1,1} = (-3\sigma_{\rm J} - \sigma_{\rm J}{\rm e}^{-i\theta} - \sigma_{\rm L} - \sigma_{\rm g}^{(1)})\varphi_{1,2},\\
    %\label{eq:Mapping_Circuit_4}
    (4):& \quad - \sigma_{\rm J}(\varphi_{n-1,n} + \varphi_{n+1,n}) - \sigma_{\rm J}{\rm e}^{i\theta}\varphi_{n,n-1} - \sigma_{\rm J}{\rm e}^{-i \theta}\varphi_{n,n+1} -\sigma_{\rm P}\varphi_{n-1,n-1} =\\
    & \hspace{5cm} = (-2\sigma_{\rm J}(1 + \cos{\theta}) -\sigma_{\rm g} - \sigma_{\rm L} - \sigma_{\rm U} -\sigma_{\rm P}) \varphi_{n,n}, \quad n \in {\rm even},\\
    %\label{eq:Mapping_Circuit_5}
    (5):& \quad - \sigma_{\rm J}(\varphi_{m+1,n} + \varphi_{m-1,n} + \varphi_{m,n-1}) - \sigma_{\rm J}{\rm e}^{-i\theta} \varphi_{m,n+1} =\\
    & \hspace{5cm} = (-3\sigma_{\rm J} - \sigma_{\rm J}{\rm e}^{i \theta} - \sigma_{\rm L} - \sigma_{\rm g}^{(2)})\varphi_{m,n}, \quad n=m-1, 2<m<N-1,\\
    %\label{eq:Mapping_Circuit_6}
    (6):& \quad - \sigma_{\rm J}(\varphi_{m,n+1} + \varphi_{m+1,n} + \varphi_{m-1,n}) - \sigma_{\rm J}{\rm e}^{i\theta} \varphi_{m,n-1} =\\
    & \hspace{5cm} = (-3\sigma_{\rm J} - \sigma_{\rm J}{\rm e}^{-i\theta} - \sigma_{\rm L} - \sigma_{\rm g}^{(1)})\varphi_{m,n}, \quad m=n-1, 2<n<N-1,\\
    %\label{eq:Mapping_Circuit_7}
    (7):& \quad - \sigma_{\rm J}(\varphi_{n+1,n} + \varphi_{n-1,n}) - \sigma_{\rm J}{\rm e}^{i \theta}\varphi_{n,n-1} - \sigma_{\rm J}{\rm e}^{-i \theta}\varphi_{n,n+1} - \sigma_{\rm P} \varphi_{n+1,n+1} =\\
    & \hspace{5cm} = (-2\sigma_{\rm J}(1 + \cos{\theta}) - \sigma_{L} - \sigma_{\rm U} - \sigma_{\rm P} - \sigma_{\rm g})\varphi_{n,n}, \quad n \in {\rm odd},\\
    %\label{eq:Mapping_Circuit_8}
    (8):& \quad - \sigma_{\rm J}\varphi_{N-1,N} - \sigma_{\rm J}{\rm e}^{i\theta}\varphi_{N,N-1} = (-2\sigma_{\rm J}(1 + \cos{\theta}) - \sigma_{L} - \sigma_{\rm U} - \sigma_{\rm P} - \sigma_{\rm g})\varphi_{N,N},\\
    %\label{eq:Mapping_Circuit_9}
    (9):& \quad - \sigma_{\rm J}(\varphi_{m,2} + \varphi_{m+1,1} + \varphi_{m-1,1}) = (-4\sigma_{\rm J} - \sigma_{\rm L})\varphi_{1,n}, 2<m<N-1,\\
    %\label{eq:Mapping_Circuit_10}
    (10):& \quad - \sigma_{\rm J}(\varphi_{N,2}+\varphi_{N-1,1}) = (-4\sigma_{\rm J} - \sigma_{\rm L})\varphi_{N,1},\\
    %\label{eq:Mapping_Circuit_11}
    (11):& \quad - \sigma_{\rm J}(\varphi_{m,n+1} + \varphi_{m,n-1} + \varphi_{m-1,n} + \varphi_{m+1,n}) = (-4\sigma_{\rm J} - \sigma_{\rm L})\varphi_{m,n}, 2<m,n<N-1, |m-n|>1,
\end{align*}
with admittances $\sigma_{\rm J}=-i2\pi f C_{\rm J}$, $\sigma_{\rm P}=-i2\pi f C_{\rm P}$, $\sigma_{\rm U}=-i2\pi f C_{\rm U}$, $\sigma_{\rm L}=i/(2\pi f L)$, $\sigma_{\rm g}^{(1)}=\sigma_{\rm J}(1-{\rm e}^{i\theta})$, $\sigma_{\rm g}^{(2)}=\sigma_{\rm J}(1-{\rm e}^{-i\theta})$, $\sigma_{g}=2\sigma_{\rm J}(1-\cos{\theta})$, and $\sigma_{\rm J}{\rm e}^{\pm\theta}$ being the admittance of elements $Z(\theta)$. Here, we assume the notation for time-varying fields $\varphi_{mn} \propto {\rm e}^{-i\omega t}$ instead of the common radio-physics notation $\varphi_{mn} \propto {\rm e}^{i\omega t}$ for consistency with the Schr{\"o}dinger-type equation describing the tight-binding model. Dividing the equations by $\sigma_{\rm J}$, we obtain a precise mapping between the two sets of equations with the following relations between the parameters of the tight-binding model and equivalent electrical circuit elements (assuming $J=1$):
\begin{align}
    & P=-\dfrac{C_{\rm P}}{C_{\rm J}}\\
    & U=\dfrac{C_{\rm P}+C_{\rm U}}{2C_{\rm J}}\\
    & \varepsilon = -\dfrac{\sigma_{\rm L}}{\sigma_{\rm J}}-4=-\dfrac{i}{2\pi f L} \dfrac{-1}{i 2\pi f C_{\rm J}} - 4= \dfrac{f_0^2}{f^2}-4
\end{align}
where $f_{0} = 1/(2 \pi \sqrt{L C_{\rm J}})$ is a characteristic frequency of the circuit.

%__________________________Simulations______________________________
\section{Numerical simulations of electric circuit}

In this Section, we describe the details of numerical simulations of the designed electrical circuits. The simulations are performed with Keysight Advanced Design System (ADS) software package. We investigate three different circuits with $15 \times 15$ nodes, corresponding to the different types of quantum statistics: $\theta=0$, $\theta\approx 1$, and $\theta=\pi$. For the chosen parameters of the tight-binding model ($J=1$, $P=-0.75$ and $U=1.5$), part of the elements in all three circuits remain the same: $L=23.21$~$\mu {\rm H}$, $C_{\rm J}=1$~$\mu {\rm F}$, $C_{\rm P}=0.75$~$\mu {\rm F}$, $C_{\rm U}=2.25$~$\mu {\rm F}$ for non-corner nodes, and $C_{\rm U}=2.91$~$\mu {\rm F}$ for the nodes with coordinates ($1,1$) and ($15,15$). Besides, we consider the following elements with $\theta$-dependent admittances for each case of quantum statistics:

\begin{enumerate}

\item Bosonic statistics, $\theta=0$. The impedance of the diagonal elements $Z$ becomes equal to the impedance of the capacitor $C_{\rm J}$. Grounding elements for the nearest to the diagonal nodes $Z_{\rm g}^{(1)}$ and $Z_{\rm g}^{(2)}$ as well as the additional grounding capacitor for diagonal nodes $C_{\rm g}$ are not required. \\

\item For emulating fractional quantum statistics, we assume $\theta=1$. In this case, the impedance of the diagonal elements $Z$ is complex. In particular, the real part corresponding to the resistance $R=15.8$~Ohm changes the sign for the different current directions (from bottom node to the top one it is positive, and from top node to the bottom one it is negative). The imaginary part is equivalent to those of the capacitor $C=0.48$~$\mu {\rm F}$. The grounding element for the left nearest to the diagonal node $Z_{\rm g}^{(1)}$ corresponds to the resistance $R_{\rm g}^{(1)}=11.77$~Ohm and capacitor $C_{\rm g}^{(1)}=2$~$\mu {\rm F}$. The right co-diagonal grounding element $Z_{\rm g}^{(2)}$ is the same combination as the left one, except the sign of the resistance: $R_{\rm g}^{(2)}=-11.77$~Ohm and $C_{\rm g}^{(2)}=2$~$\mu {\rm F}$. The implementation of negative resistance is discussed below. The additional grounding capacitor is $C_{\rm g}=1$~$\mu {\rm F}$. Note also that positive and negative resistances are balanced, which makes the circuit Hermitian in agreement with the Hermitian nature of the original tight-binding model.\\

\item Pseudo-fermionic statistics, $\theta=\pi$. In such a case, the diagonal element $Z$ contains inductance $L_{\rm J}=190.21~\mu{\rm H}$. The grounding elements of the co-diagonal nodes are represented by the capacitors for each side from the diagonal $C_{\rm g}^{(1)}=C_{\rm g}^{(1)}=2~\mu{\rm F}$, while the additional grounding capacitors are $C_{\rm g}=4~\mu{\rm F}$.\\

\end{enumerate}

Note that in the limiting cases of statistical exchange angle $\theta=0,\pi$ the circuits consist of purely passive elements, while the circuit with intermediate value of $\theta$ requires active elements to implement negative resistance.

\begin{figure}[tbp]
    \centering
    \includegraphics[width=10cm]{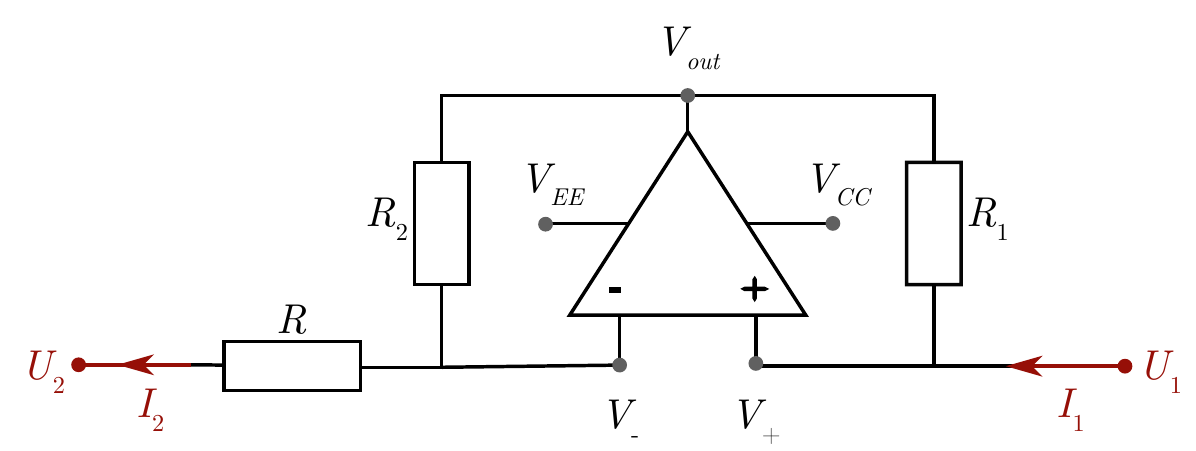}
    \caption{Negative impedance converter configuration with current inversion. The schematic includes ideal operational amplifier (OpAmp), and resistors $R$, $R_1$, and $R_2$. $V_{CC}$ and $V_{EE}$ are positive and negative supply voltages, respectively.}
    \label{fig:NIC}
\end{figure}

%__________NIC______________
To implement negative resistance, we consider a negative impedance converter (NIC) with current inversion illustrated in Fig.~{\ref{fig:NIC}}.

The input impedance of the considered schematics is defined by the equation $Z_{in}=U_1/I_1$, where $U_1$ is the input voltage and $I_1$ is the input current. As the idealized operational amplifier is considered in the simulations, its internal resistance is infinite, and no currents flow inside. Consequently, voltages of the non-inverting and inverting OpAmp's inputs and the input voltage are equal $V_+=V_-=U_1$, hence $U_1=U_2=U$. Next, we examine OpAmp's output voltage $V_{out}=I_2 R_2+U$ and establish the input current:
\begin{equation}
    I_1=\dfrac{U-V_{out}}{R_1}=\dfrac{U-I_2 R_2 -U}{R_1}=-\dfrac{I_2 R_2}{R_1}
\end{equation}
This yields the following expression for the input impedance:
\begin{equation}
    Z_{in}=\dfrac{U}{I_1}=-\dfrac{U R_1}{I_2 R_2}=-R\dfrac{R_1}{R_2}
\end{equation}

If $R_1=R_2$, the input impedance is $Z_{in}=-R$. Then, if we consider the same system but supply test voltage source to the left side, the input impedance equals resistance $R$ with no sign inversion. For the simulation of such electrical circuit, we use ideal OpAmps with $1$~GHz bandwidth, and supply voltages $\pm 5$~V. As the admittance of diagonal elements $Z(\theta)$ for the case $\theta=1$ also has the imaginary part corresponding to the impedance of a capacitor, we use a parallel connection of a NIC and a capacitor.

Finally, we perform an S-Parameter simulation to obtain the magnitude of the input impedance for each node in the frequency range $8-16$~kHz with a $10$~Hz step. Within this simulation, we set Q-factor of inductors $L$ and $L_{\rm J}$ at the value of $Q_{\rm L}=200$. Such values of Q-factor for inductors in kHz frequency range can be achieved either by (i) using coils made of a litz wire that consists of a large number of thin wire strands and allows to reduce the skin and proximity effects; (ii) using materials with a very high permeability as an inductors' core; (iii) cooling inductive coils down to cryogenic temperatures with liquid nitrogen to reduce electron-phonon scattering and increase the conductivity of a wires' material; and (iv) optimizing the coil shape~\cite{2020_Rikhter}. Moreover, when realizing the proposed circuit experimentally, lower Q-factors of inductors and operational amplifiers' bandwidths can be applied at the cost of less pronounced bandgap in the circuit spectrum and lower quality of the observed topological states.

%__________________________References_______________________________

%\bibliography{Anyons_Supplementary_BibFile}

%